\documentclass[prl,aps,twocolumn,floatfix,showpacs]{revtex4}
\usepackage{bm,graphicx,epsf}

\newcommand{\be}{\begin{equation}}
\newcommand{\ee}{\end{equation}}
\newcommand{\bea}{\begin{eqnarray}}
\newcommand{\eea}{\end{eqnarray}}
\newcommand{\phrl}[1]{Phys.~Rev.~Lett. {\bf #1}}
\newcommand{\phrb}[1]{Phys.~Rev.~B {\bf #1}}

\newcommand{\bib}{\bibitem}
\newcommand{\lb}{\left[}
\newcommand{\rb}{\right]}
\newcommand{\lp}{\left(}
\newcommand{\rp}{\right)}

\begin{document}

\title{Electronic structure of a single vortex in $d$-wave superconductor
        revisited}
\author{Sanjay Gupta}
\affiliation{$^{*}$Theoretical Physics Division, Indian Association for the  
Cultivation of Sciences, Kolkata, India}

\date{\today}


\begin{abstract}
The present work deals with the study of $d$-wave superconductor in presence
of a single vortex placed at the centre of the 2D lattice using the $t-t'-J$ 
model within the renormalized mean field theory. It is found that the in
 absence of the vortex the ground state has a d-wave configuration. In
 presence of the vortex, the superconducting order parameter, above the
 critical doping, drops to a low non zero value within a few lattice points
 from the vortex (that is within the the vortex core) and beyond it converges
 to the constant value in absence of vortex. We observe that above the
 critical doping things are consistent with the experimental results while
 there is an anomalous rise in the superconducting order parameter at very
 low doping within the vortex core. This we feel is because of
 antiferromagnetic ordering taking place within the vortex core at low doping.
 
\end{abstract}

\maketitle


{\it Introduction:} In a type-II superconductor, magnetic field enters into
the system in terms of vortices. The order parameter owing to broken
symmetry in a superconductor aquires a topological phase of $2\pi$
for an winding. The structure of the
vortices in superconductors is traditionally an interesting subject of 
research in condensed matter physics.
In a conventional BCS superconductor with $s$-wave pairing symmetry
 \cite{Caroli}, 
the formation of quasiparticle bound states inside the core of a vortex 
that had been predicted long back, has been verified in scanning tunneling
microscopy (STM) measurement \cite{Hess,Gygi,Aprile,Pan}.

But a similar convincing picture in an unconventional superconductor
like cuprate superconductor with
$d_{x^2-y^2}$-wave pairing is still elusive. This is because
the pair potential in the latter kind of superconductor is nonlocal and 
it has nodes. 
The conventional weak coupling BCS theory for $d$-wave vortices predicts
zero energy peak in the local density of states at the vortex core for the
Bogoliubov quasiparticles from the solution of the Bogoliubov-de Gennes (BdG) 
equations \cite{Wang}. However, the scanning tunneling microscopy
measurements on the cuprate superconductors seem to reveal a bound
state inside the vortex core at fairly high energy \cite{Aprile,Pan}. This fact is counter intuitive, since quasiparticle excitation is possible even at
 zero energy in a pure $d$-wave superconductor \cite{Kallin, Ghosal1}. 
There have been several proposals like  $d_{x^2-y^2}+id_{xy}$ \cite{Lee}
 state,a staggered flux state, and an antiferromagnetic vortex core \cite{Ghosal2}to describe the absence
 of ZEP inside the core. Subsequently, neutron scattering experiments near
 optimal doping \cite{Lake1}, $\mu$SR and NMR experiments in underdoped
 compounds \cite{Lake2} seem to
 suggest the existence of the enhancement of the AF correlation inside the core
\cite{Lee, Franz}. Therefore the absence of zero bias tunneling conductance
 in the STM
 measurements may be due to the AF gap developed inside the core. In the
present work we study the d-wave superconductivity using the $t-t'-J$ model 
using renormalized mean field theory(RMFT). In this scheme we use Gutzwiller
 projected wavefunction which renormalizes the parameters of the Hamiltonian.
 The resulting Hamiltonian is decoupled using the
 Unrestricted Hartree-Fock scheme and then the corresponding Bogoliubov-DeGennes
 equations are solved in a self consistent fashion. The central idea is to
capture the strong electronic correlation efficiently. We study the
simplest case to begin with where decoupling is done only in favour
of the superconducting order parameter. Depending on the results of this
 study we intend to include the decoupling in favour of antiferromagnetic
 order in future in the required regime of doping. 

\vskip .1in
{\it The Model:}
 We consider a Hamiltonian 
\bea
H &=& -t\sum_{i,\delta_1,\sigma}c^\dagger_{i\sigma}c_{i+\delta_1\sigma}
 +t'\sum_{i,\delta_2,\sigma}c^\dagger_{i\sigma}c_{i+\delta_2\sigma} 
  -\mu \sum_{i,\sigma} c^\dagger_{i\sigma}c_{i\sigma} \nonumber\\
& & + J\sum_{i,\delta_1} \lp \bm{S}_i\bm{\cdot}\bm{S}_{i+\delta_1}
          -\frac{1}{4}n_in_{i+\delta_1} \rp
\label{Hamiltonian}
\eea
with no doubly-occupied sites.
Here $c^\dagger_{i\sigma}$ represents creation of an electron with spin
$\sigma = \uparrow$ or $\downarrow$ at site $i$, $\delta_1 = \pm \hat{x},\,
\pm \hat{y}$, $\delta_2 = \pm \hat{x} \pm \hat{y}$. (Here the inter-site
distance is chosen to be unity.) The hopping energies $t$ and $t'$  are
for the nearest neighbor (NN) and next nearest neighbour (NNN) 
respectively, and $\mu$ is the chemical potential 
which fixes the average density of electrons $n=1-x$ with hole doping $x$.
The spin operator $S^a_i = \frac{1}{2}\sum_{\alpha\beta}
     c_{i\alpha}^\dagger \sigma^a_{\alpha\beta} c_{i\beta}$ 
with $\sigma^a$ being the Pauli matrices, and
the density opeartor $n_i = c_{i\uparrow}^\dagger c_{i\uparrow}
 +c_{i\downarrow}^\dagger c_{i\downarrow}$.
The exchange energy $J$ term in Eq.~(\ref{Hamiltonian}) may be decomposed
into NN bond pairing amplitude
$\frac{J}{2}\lb \langle c_{i\uparrow}
     c_{i+\delta_1\downarrow} \rangle_0  - \langle
c_{i\downarrow} c_{i+\delta_1 \uparrow} \rangle_0 \rb $
which describes $d_{x^2-y^2}$-superconductor with a mean field state
$\vert \Phi_0\rangle$. The suppression of double occupancy is achieved
by the Gutzwiller projection of $\vert \Phi_0\rangle$ into
the state $\vert \Phi\rangle = \Pi_i(1-n_{i\uparrow}n_{i\downarrow})
\vert \Phi_0\rangle$. The expectation values in this new state is related
with that of ordinary mean field state as follows: 
$\langle \Phi\vert c_{i\sigma}^\dagger c_{j\sigma} \vert \Phi\rangle
= g_t \langle \Phi_0\vert c_{i\sigma}^\dagger c_{j\sigma} \vert \Phi_0\rangle$
and $\langle \Phi\vert \bm{S}_i\cdot \bm{S}_{j} \vert \Phi\rangle 
= g_s \langle \Phi_0\vert \bm{S}_i\cdot \bm{S}_{j} \vert \Phi_0\rangle$, where
$g_t = 2x/(1+x)$ and $g_s = 4/(1+x)^2$ \cite{Anderson,Zhang}.
Therefore the Guzwiller renormalized mean field Hamiltonian
for a superconductor becomes
\bea
{\cal H} &=& -\tilde{t}\sum_{i,\delta_1,\sigma}c^\dagger_{i\sigma}c_{i+\delta_1\sigma}
 +\tilde{t}'\sum_{i,\delta_2,\sigma}c^\dagger_{i\sigma}c_{i+\delta_2\sigma} 
  -\tilde{\mu} \sum_{i,\sigma} c^\dagger_{i\sigma}c_{i\sigma} \nonumber\\
  & & +\sum_{i,\delta_1}\lb \Delta_{i,i+\delta_1} \lp 
       c^\dagger_{i\uparrow} c^\dagger_{i+\delta_1\downarrow} -
     c^\dagger_{i\downarrow} c^\dagger_{i+\delta_1\uparrow}  \rp +h.c. \rb \, ,
\label{Hamiltonian_eff}
\eea
where $\tilde{t} = g_t t$, $\tilde{t}'=g_t t'$, 
$\tilde{\mu}$ is the renormalized chemical potential which fixes $x$
for ${\cal H}$ and the effective pair amplitude
\be
\Delta_{i,i+\delta_1} = \frac{\tilde{J}}{2}\lb \langle c_{i\uparrow}
     c_{i+\delta_1\downarrow} \rangle  - \langle
c_{i\downarrow} c_{i+\delta_1 \uparrow} \rangle \rb 
\label{delta_x}
\ee
with $\tilde{J} = (3g_s +1)/4$.
The Hamiltonian (\ref{Hamiltonian_eff}) can be diagonalized 
using standard Bogoliubov
transformatin leading to the usual BdG equations
for both kind of Bogoliubov quasiparticles:
\bea
-\tilde{t}\sum_{\delta_1}u^n_{i+\delta_1}+ 
\tilde{t}'\sum_{\delta_2}u^n_{i+\delta_2}
+\sum_{\delta_1} \Delta_{i,i+\delta_1}v^n_{i+\delta_1}
   = (E_n+\tilde{\mu})u^n_i & & 
\label{BDG1a} \\
\tilde{t}\sum_{\delta_1}v^n_{i+\delta_1}- 
\tilde{t}'\sum_{\delta_2}v^n_{i+\delta_2}
+\sum_{\delta_1} \Delta^{*}_{i,i+\delta_1}u^n_{i+\delta_1}
   = (E_n-\tilde{\mu})v^n_i & &
\label{BDG1b}
\eea
where $E_n$ is the energy for the $n$-th eigen value, and $u^n_i$ and
$v^n_i$ are the corresponding amplitudes at the $i$-th site for creation
of a spin-$\uparrow$ electron and destruction of a spin-$\downarrow$ hole
respectively.
The NN pairing potential can be obtained through the Gutzwiller meanfield
equation as 
\be
\Delta_{i,i+\delta_1} = \tilde{J} \sum_{n (E_n>0)}\lp u^n_i v^{*n}_{i+\delta_1} +  u^n_{i+\delta_1}v^{*n}_i \rp \, .
\label{Deltad}
\ee
For a pure $d_{x^2-y^2}$ superconductor, $\Delta_{i,i\pm \hat{x}} =
-\Delta_{i,i\pm \hat{y}} = \Delta_0$, a constant.
If we introduce a vortex,
pair amplitude aquires a phase in the center of mass coordinate (CMC)
of the pair:
$\Delta_{i,i\pm \hat{x}} = \Delta_0 e^{i\theta_{i,i\pm \hat{x}}}$ and
$\Delta_{i,i\pm \hat{y}} = -\Delta_0 e^{i\theta_{i,i\pm \hat{y}}}$
where $\theta_{i,j}$ is the angle of the CMC of
$i$ and $j$ with respect to the center of the vortex. This implies
phase winding of $2\pi$ around the vortex. Choosing realistic parameters
as $t' = t/4$ and $J=t/3$, we then solve the BDG equations 
for eigenvalues $E_n$ and eigenfunctions $(u_n^i, v_n^i)$ for different values
of $\tilde{mu}$ in the presence of a vortex placed at the center of 
an $N\times N$ ($N$ even) lattice. Even $N$ is chosen for placing the vortex 
in the center of an atomic lattice as it is the most symmetric and energetically
favorable position. We begin with the above form of $\Delta_{i,i+\delta_1}$
for a vortex and solve the BDG equations in the open bounday condition
together with self-consistency for the average
hole concentration $x= \frac{1}{N^2}\sum_i(1-2\sum_n \vert v_n^i \vert^2)$
and the average NN pair amplitude $\Delta =\frac{1}{2N(N-1)}
\sum_{i,\delta_1}\Delta_{i,i+\delta_1}$. 
\begin{figure}
\vspace{2.4cm}
\centerline{\hspace{0cm}{\epsfysize=5cm\epsfbox{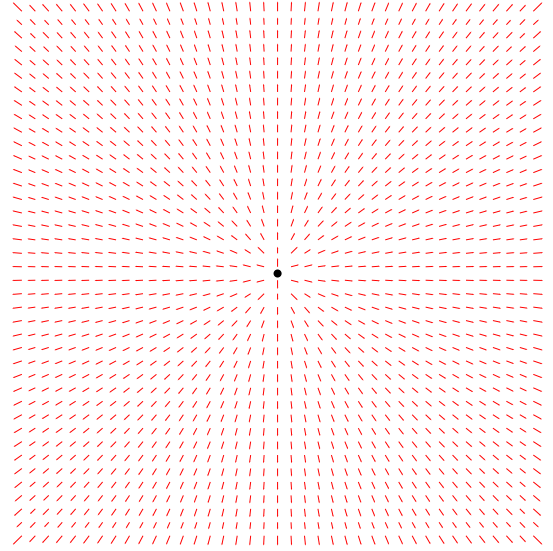}}}
\caption{Configuration of a vortex placed at the center of a $40\times 40$
 lattice. Black dot represents the position of the centre of the vortex.
 Length and angle of an arrow represent the magnitude and phase of the pair
 amplitude respectively at the centre, which is denoted by the tail of the
 arrow, of two neighbouring lattice site.}
\label{fig1}
\end{figure}
We have $N=40$, 50, and 60 in our finite size computation. However, there
is no boundary effect on the phase as has been shown in Fig.~\ref{fig1} for
 N=40.
There is no boundary twist on the phase. The bulk magnitude of pair
amplitude is independent of $N$ for the chosen numbers. 
This is demonstrated in the inset of Fig.~\ref{fig2}. 
We therefore believe that we have achieved the thermodynamic limit to predict
the electronic structure of a single vortex in high $T_c$ superconductors.

\vspace{0.5cm}
\begin{figure}
\centerline{\hspace{-1cm}{\epsfysize=6cm\epsfbox{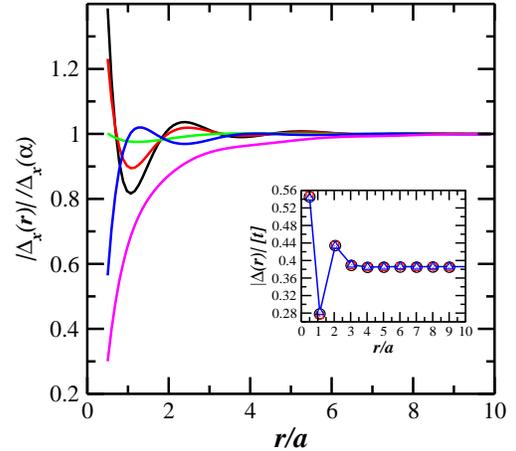}}}
\caption{The behavior of $\Delta (r)$ in the unit of its value at the bulk as a function of the distance $r$ away from the center of the vortex for different
 values of $x$. The values of $x$ are 0.05, 0.08, 0.13, 0.31, and 0.49 from top to bottom at the left edge of the curves. The inset shows $\Delta (r)$ when 
 $y=0.5$ for N= 40 (circle),  50 (square), and 60 (triangle) at the hole
 concentration $x \approx 0.05$}
\label{fig2}
\end{figure}


\begin{figure}
\vspace{-1.0cm}
\centerline{\hspace{-1cm}{\epsfysize=12cm\epsfbox{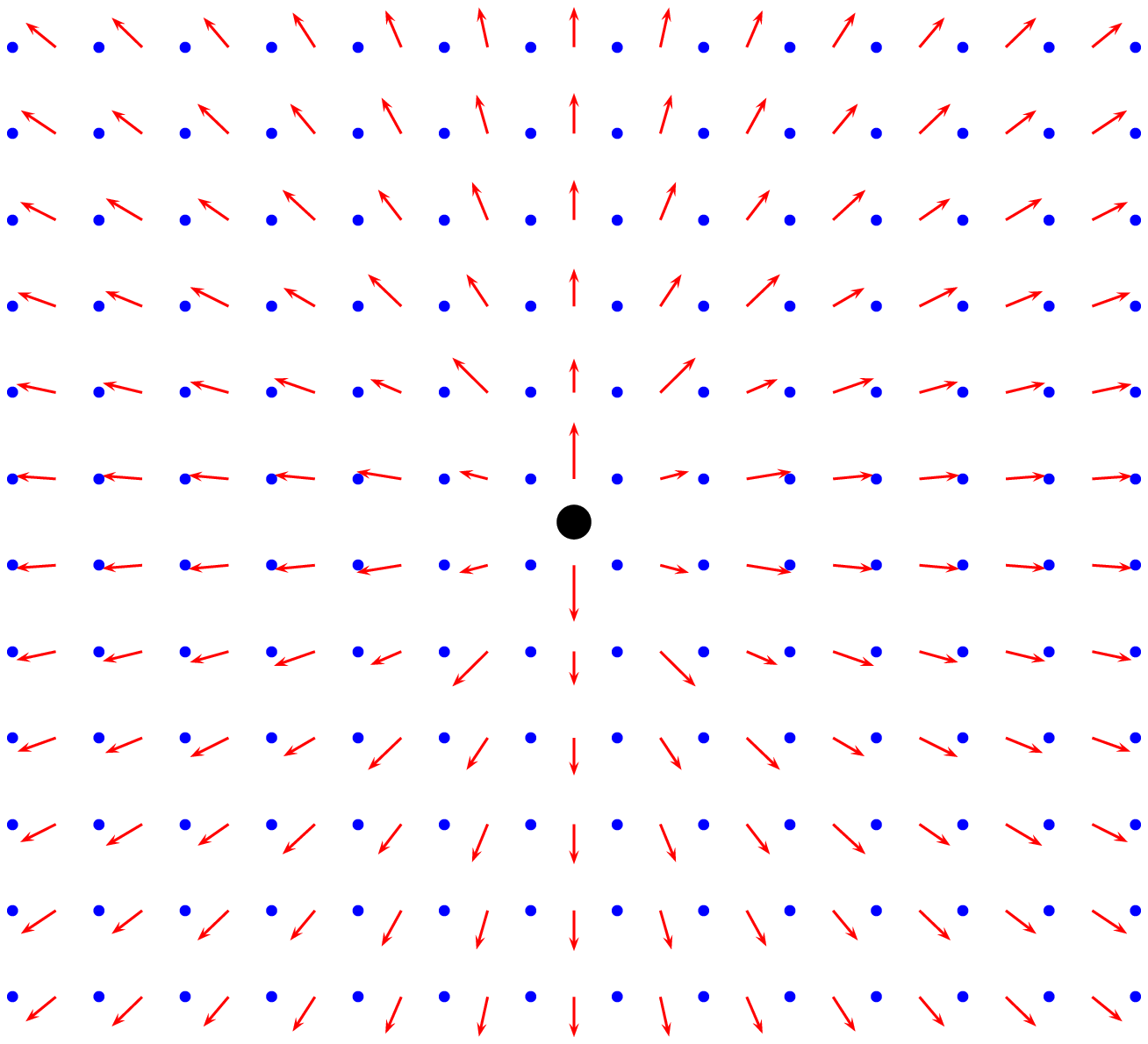}}}
\vspace{-6.5cm}
\centerline{\hspace{-1cm}{\epsfysize=12cm\epsfbox{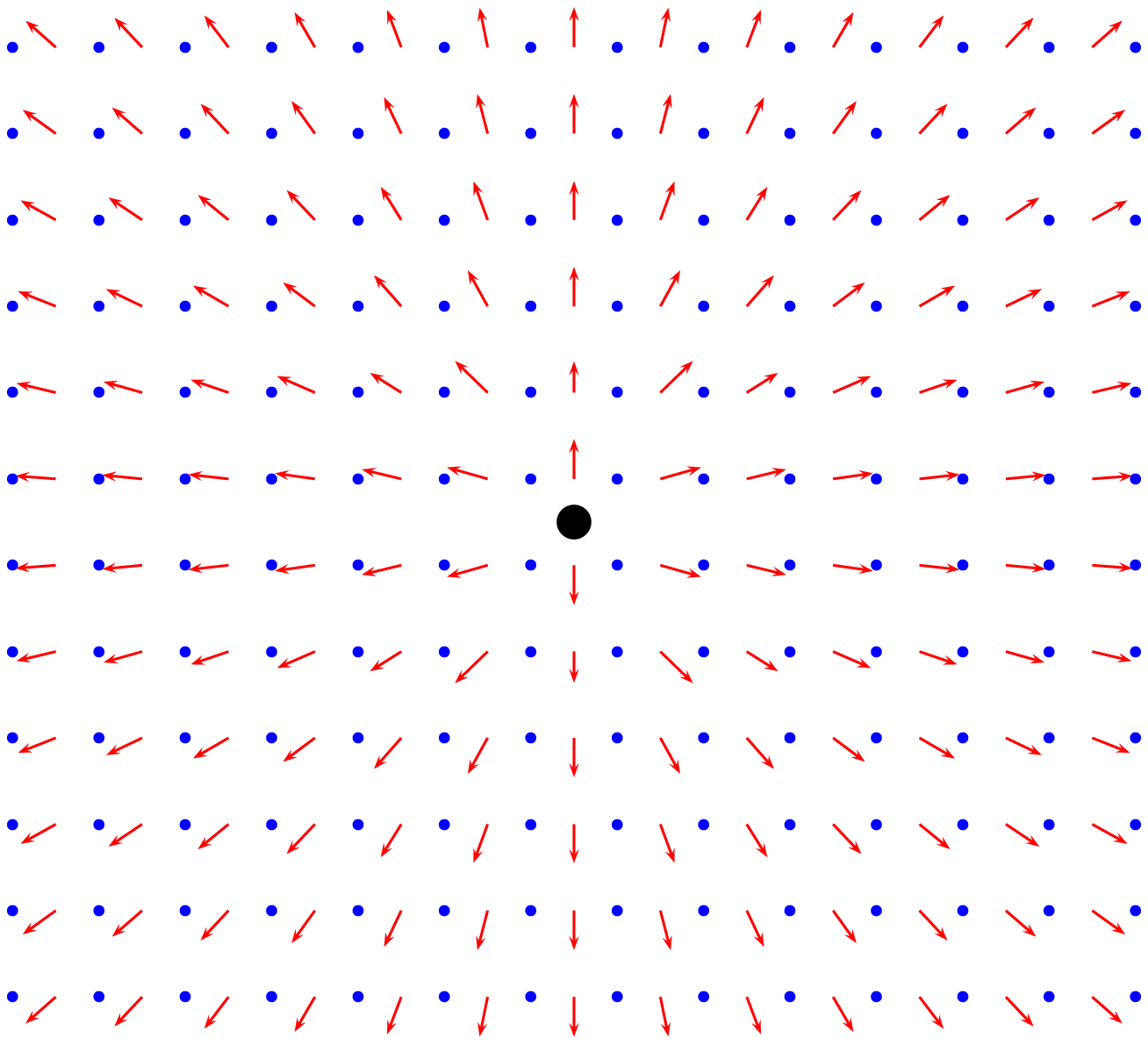}}}
\vspace{-6.5cm}
\centerline{\hspace{-1cm}{\epsfysize=12cm\epsfbox{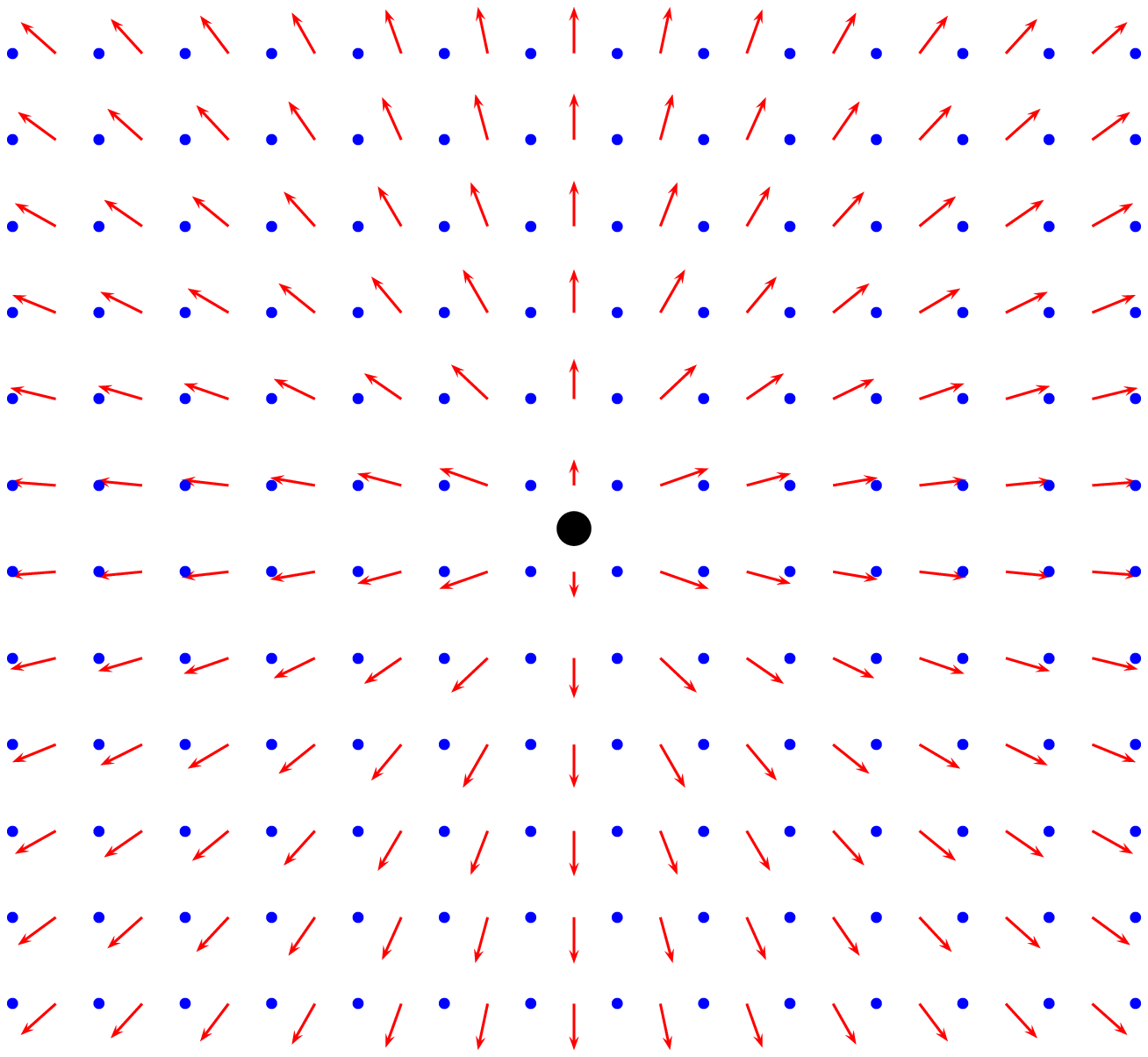}}}
\vspace{-6.5cm}
\centerline{\hspace{-1cm}{\epsfysize=12cm\epsfbox{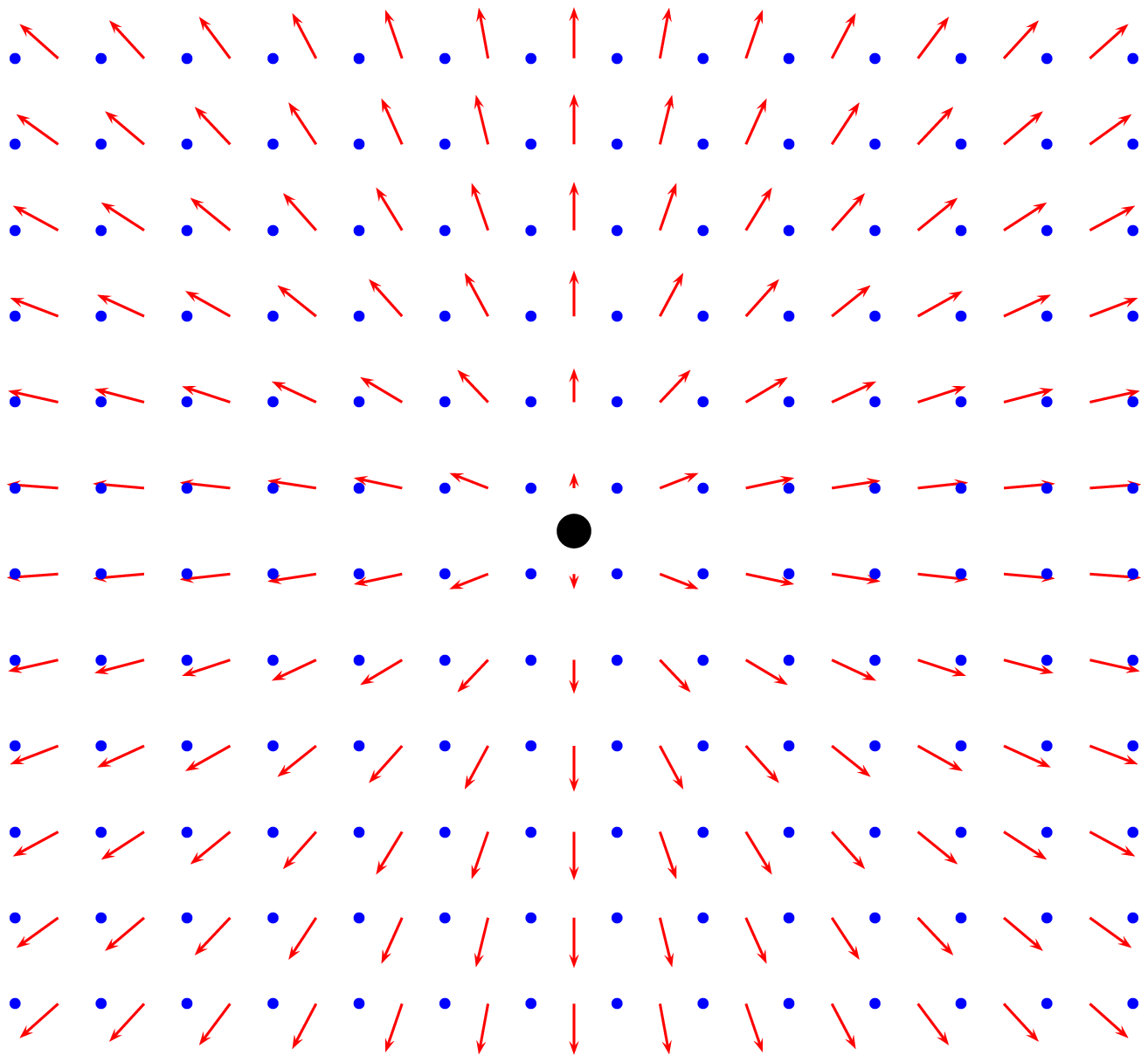}}}
\vspace{-5.5cm}
\caption{Structure of a single vortex in a $60\times60$ lattice with hole
 concentrations $x= 0.05$, 0.13, 0.31, and 0.49 (from top to bottom
 panels). Black dot represents the center of the vortex. Blue dots
 are the position of the $Cu$ atoms in the lattice. Length, direction,
 and tail of the arrows represent the magnitude of bond-pair amplitude,
 phase of bond-pair amplitude, and the center of the bonds respectively.}
\label{fig3}
\end{figure}

\begin{figure}
\centerline{\hspace{1cm}{\epsfysize=4cm\epsfbox{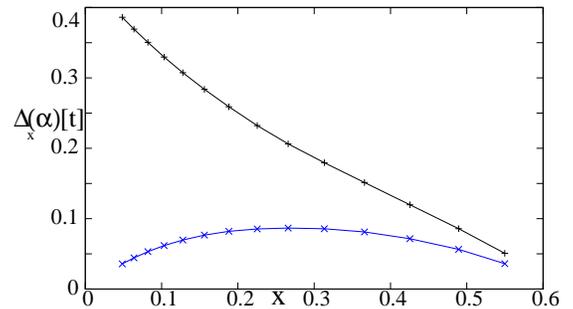}}}
\caption{The bulk values of pair-amplitude as a function of $x$
 (upper-curve). Its renormalization with the factor $g_t = 2x/(1+x)$
 is shown in the lower-curve.}
\label{fig4}
\end{figure}

For the largest system with $N=60$ that we have considered, the
 magnitude and phase of the pair amplitude between the electrons in
 the bonds around a single vortex is shown in Fig.~\ref{fig3} for the
 hole concentrations 0.05, 0.13, 0.31, and 0.49. The phase winding around
 the vortex in each case is $2\pi$. The magnitude of the pair amplitude
 $\Delta (r)$ is constant away from the core of the vortex, i.e.,
 $\Delta (r\to \infty) = \Delta_0$. Its dependence on $x$ is shown in
 Fig.~\ref{fig4}. However, the behavior of $\Delta (r)$ inside
 the core have several features depending on $x$. (i) At the highly underdoped
regime, unlike any other superconductor, $\Delta (r)$ is greater than
 $\Delta_0$ at the closest bond from the center of the vortex, it is lesser
 than $\Delta_0$ in the immediate next bond and the next bond onwards it is
 almost close to $\Delta_0$. (ii) Near quantum critical point, $x=1/8$,
 $\Delta (r)$ is almost independent of $r$. (iii) Near optimal doping, the
 behavior of $\Delta (r)$ inside the core is exactly opposite to the case of
 highly underdoped regime. (iv) At the highly overdoped regime, its
 behaviour near the center of the vortex is the same as conventional
 vortex, i.e., its value gradually decreases towards the center of the 
vortex. Since the calculation is performed in a lattice, the value of
 $\Delta (r)$ will not be the exactly same for same values of $r$ but
 at different bonds. We therefore perform best fit for $\Delta (r)$ and
 have demonstrated its behaviour as a function of $r$ for different values
 of $x$ in Fig.~\ref{fig2}. $\Delta (r)$ is scaled by $\Delta_0$ for
 different values of $x$. The following remarkable features are found.
 (i)The structure of the core at highly overdoped regime is conventional,
 i.e., the pair amplitude approaches zero towards the center of the vortex
 and then smoothly attains its asymptotic value. (ii) In highly underdoped
 regime, the pair amplitude has larger value at deep inside the core
 compared to the same in the bulk and it oscillates about the bulk value
 before attaining the asymptotic value. (ii) The regime around optimal
 doping shows that has smaller value (but does not converge to zero as
 $r\to 0$) at deep inside the core compared to the same in the bulk and
 it oscillates about the bulk value before attaining the asymptotic value.
 (iv) Near quantum critical point in the underdoped regime, pair amplitude
 remains almost constant everywhere. From the curves of Fig.~2, we estimate the coherence length to be about $3$ times
 atomic lattice constant. This is consistent with the experiments.
Figure~\ref{fig4} depicts the behaviour of the absolute value of the 
$d$-wave order parameter away from the vortex core as a function of the
 doping $x$ with (lower curve) and without(upper  curve) the renormalizing
 factor being multiplied with it. We observe that the dome like feature in
 the $d$-wave  order parameter is obtained with the renormalizing factor
 $g_{t} = 2x/(1+x)$ being multiplied as shown in the lower curve. This is
again at par with the experimental results. So, as far as the bulk 
 behaviour of the $d$-wave parameter is concerned the present
scheme of decoupling the Hamiltonian only in favour of the superconducting
order parameter is sufficient. The antiferromagnetic ordering, we feel is
important for the underdoped regime to explain the physics inside the
vortex core. The quasiparticle local density of states (DOS) is given by:

\be
\rho_i(\epsilon)=\sum_n\left[\vert u^n_i\vert^2\delta(\epsilon-E_n)+\vert
 v^n_i\vert^2\delta(\epsilon+E_n)\right]
\ee

\begin{figure}
\centerline{\hspace{1cm}{\epsfysize=5cm\epsfbox{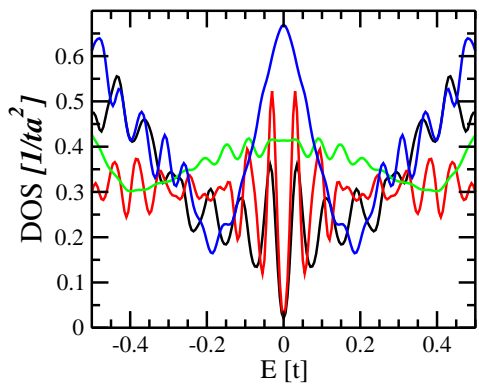}}}
\caption{Quasiparticle density of states as a function of quasiparticle energy
 for $x=0.05$, 0.13, 0.31, and 0.49 respectively from bottom to top seen at
 $E=0$.}
\label{fig5}
\end{figure}

Figure~\ref{fig5} show the DOS at one of the four nearest atomic site from
 the center of the vortex for N=60, and $x=0.05$, 0.13, 0.31 and 0.49. For
 highly overdoped regime, the DOS at the core has large ZEP as was found by
 the conventional study. Near optimally doped regime, the DOS is large but
 does not show any peak near zero energy. The underdoped regime shows that
 the DOS is vanishingly small at zero energy but shows a peak at
 $E \approx 0.03t$. This peak is consistent with the large tunneling current
 observed at fairly high energy.

{\bf Conclusion}
\vskip .1in
The present work reveals that above the critical doping, decoupling the
Hamiltonian only in favour of the superconducting
 order parameter is giving results consistent with the
 experiments for both the bulk and the vortex core. Below critical
 doping the behaviour of the d-wave order parameter in the bulk 
 is consistent with the experimental findings but inside the vortex core it
shows anomalous behaviour as it rises
 sharply on approaching the vortex centre.
 We believe that in the underdoped regime the antiferromagnetic ordering is
 important to get the correct physics inside the vortex core. Hence decoupling
 the Hamiltonian both in favour of the antiferromagnetic and superconducting
order parameter and then solving it self consistently would give results at
 par with the experiments in the vortex core. 
\vskip .1in
Acknowledgement: The author is grateful to Prof Sudhansu Mandal of
 Indian Association For the Cultivation of Sciences, Kolkata for
 suggesting the problem and for some important discussions. The author
also thanks Tribikram Gupta for the help he provided in drawing one
 of the figures.
\vskip .1in
*Present Address: Department of Physics, Indian Institute of Technology,
New Delhi.

\end{document}